\documentclass[twocolumn]{aastex631}

\newcommand\tess{\textit{TESS}}
\newcommand{\kms}{\ensuremath{\rm km\,s^{-1}}}
\newcommand{\ms}{\ensuremath{\rm m\,s^{-1}}}

\shorttitle{TOI-4600 b \& c}
\shortauthors{Mireles et al.}
\graphicspath{{./}{}}

\begin{document}

\title{TOI-4600 b and c: Two long-period giant planets orbiting an early K dwarf}

\author[0000-0002-4510-2268]{Ismael Mireles}
\affiliation{Department of Physics and Astronomy, The University of New Mexico, Albuquerque, NM 87106, USA}

\author{Diana Dragomir}
\affiliation{Department of Physics and Astronomy, The University of New Mexico, Albuquerque, NM 87106, USA}

\author[0000-0002-4047-4724]{Hugh P. Osborn}
\affiliation{NCCR/PlanetS, Physikalisches Institut, University of Bern, Gesellsschaftstrasse 6, 3012 Bern, Switzerland}
\affiliation{Department of Physics and Kavli Institute for Astrophysics and Space Research, Massachusetts Institute of Technology, Cambridge,
MA 02139, USA}

\author[0000-0002-2135-9018]{Katharine Hesse}
\affiliation{Department of Physics and Kavli Institute for Astrophysics and Space Research, Massachusetts Institute of Technology, Cambridge,
MA 02139, USA}

\author[0000-0001-6588-9574]{Karen A.\ Collins}
\affiliation{Center for Astrophysics \textbar \ Harvard \& Smithsonian, 60 Garden Street, Cambridge, MA 02138, USA}

\author{Steven Villanueva}
\affiliation{NASA Goddard Space Flight Center
8800 Greenbelt Rd, Greenbelt, MD 20771, USA}

\author{Allyson Bieryla}
\affiliation{Center for Astrophysics \textbar \ Harvard \& Smithsonian, 60 Garden Street, Cambridge, MA 02138, USA}

\author[0000-0002-5741-3047]{David R. Ciardi}
\affiliation{NASA Exoplanet Science Institute-Caltech/IPAC, Pasadena, CA 91125, USA}

\author[0000-0002-3481-9052]{Keivan G.\ Stassun}
\affiliation{Department of Physics and Astronomy, Vanderbilt University, Nashville, TN 37235, USA}

\author{Mallory Harris}
\affiliation{Department of Physics and Astronomy, The University of New Mexico, Albuquerque, NM 87106, USA}

\author[0000-0001-6513-1659]{Jack J. Lissauer}
\affiliation{NASA Ames Research Center, Moffett Field, CA 94035, USA}

\author[0000-0001-8227-1020]{Richard P. Schwarz}
\affiliation{Center for Astrophysics \textbar \ Harvard \& Smithsonian, 60 Garden Street, Cambridge, MA 02138, USA}

\author{Gregor Srdoc}
\affiliation{Kotizarovci Observatory, Sarsoni 90, 51216 Viskovo, Croatia}

\author[0000-0003-1464-9276]{Khalid Barkaoui}
\affiliation{Astrobiology Research Unit, Universit\'{e} de Li\`{e}ge, 19C All\'{e}e du 6 Ao\^{u}t, 4000 Li\'{e}ge, Belgium}
\affiliation{Department of Earth, Atmospheric and Planetary Science, Massachusetts Institute of Technology, 77 Massachusetts Avenue, Cambridge, MA 02139, USA}
\affiliation{Instituto de Astrof\'isica de Canarias (IAC), Calle V\'ia L\'actea s/n, 38200, La Laguna, Tenerife, Spain}

\author{Arno Riffeser}
\affiliation{University-Observatory, Ludwig-Maximilians-University, Scheinerstrasse 1, D-81679 Munich, Germany}

\author[0000-0001-9504-1486]{Kim K. McLeod}
\affiliation{Wellesley College, Wellesley, MA, USA}

\author{Joshua Pepper}
\affiliation{Department of Physics, Lehigh University, 16 Memorial Drive East, Bethlehem, PA 18015, USA}

\author[0000-0001-8105-0373]{Nolan Grieves}
\affiliation{Observatoire astronomique de l\'{}Universit\'{e} de Gen\`{e}ve, 51 Chemin des Maillettes, 1290 Versoix, Switzerland}

\author{Vera Maria Passegger}
\affiliation{Instituto de Astrof\'{i}sica de Canarias, c/ V\'{i}a L\'{a}ctea s/n, 38205 La Laguna, Tenerife, Spain}
\affiliation{Departamento de Astrof\'{i}sica, Universidad de La Laguna, 38206 La Laguna, Tenerife, Spain}
\affiliation{Hamburger Sternwarte, Universität Hamburg, Gojenbergsweg 112, 21029 Hamburg, Germany}
\affiliation{Homer L. Dodge Department of Physics and Astronomy, University of Oklahoma, 440 West Brooks Street, Norman, OK 73019, USA}

\author[0000-0003-2417-7006]{Sol\`{e}ne Ulmer-Moll}
\affiliation{Observatoire astronomique de l\'{}Universit\'{e} de Gen\`{e}ve, 51 Chemin des Maillettes, 1290 Versoix, Switzerland}

\author[0000-0001-8812-0565]{Joseph E. Rodriguez}
\affiliation{Center for Data Intensive and Time Domain Astronomy, Department of Physics and Astronomy, Michigan State University, East Lansing, MI, 48824, USA}

\author[0000-0002-2457-7889]{Dax L. Feliz}
\affiliation{American Museum of Natural History, 200 Central Park West, Manhattan, NY 10024, USA}

\author[0000-0002-8964-8377]{Samuel Quinn}
\affiliation{Center for Astrophysics \textbar \ Harvard \& Smithsonian, 60 Garden Street, Cambridge, MA 02138, USA}

\author[0000-0001-6037-2971]{Andrew W. Boyle}
\affiliation{NASA Exoplanet Science Institute-Caltech/IPAC, Pasadena, CA 91125, USA}

\author[0000-0002-9113-7162]{Michael Fausnaugh}
\affiliation{Department of Physics and Kavli Institute for Astrophysics and Space Research, Massachusetts Institute of Technology, Cambridge, MA 02139, USA}


\author[0000-0001-9269-8060]{Michelle Kunimoto}
\affiliation{Department of Physics and Kavli Institute for Astrophysics and Space Research, Massachusetts Institute of Technology, Cambridge, MA 02139, USA}

\author[0000-0002-4829-7101]{Pamela Rowden}
\affiliation{Royal Astronomical Society, Burlington House, Piccadilly, London W1J 0BQ, UK}

\author[0000-0001-7246-5438]{Andrew Vanderburg}
\affiliation{Department of Physics and Kavli Institute for Astrophysics and Space Research, Massachusetts Institute of Technology, Cambridge, MA 02139, USA}

\author[0000-0002-4715-9460]{Bill Wohler}
\affiliation{NASA Ames Research Center, Moffett Field, CA 94035, USA}
\affiliation{SETI Institute, Mountain View, CA 94043, USA}

\author[0000-0002-4715-9460]{Jon~M.~Jenkins}
\affiliation{NASA Ames Research Center, Moffett Field, CA 94035, USA}

\author[0000-0001-9911-7388]{David~W.~Latham}
\affiliation{Harvard-Smithsonian Center for Astrophysics, 60 Garden St, Cambridge, MA 02138, USA}

\author[0000-0003-2058-6662]{George R. Ricker}
\affiliation{Department of Physics and Kavli Institute for Astrophysics and Space Research, Massachusetts Institute of Technology, Cambridge, MA 02139, USA}

\author[0000-0002-6892-6948]{Sara Seager}
\affiliation{Department of Earth, Atmospheric, and Planetary Sciences, Massachusetts Institute of Technology, Cambridge, MA 02139, USA}
\affiliation{Department of Physics and Kavli Institute for Astrophysics and Space Research, Massachusetts Institute of Technology, Cambridge, MA 02139, USA}
\affiliation{Department of Aeronautics and Astronautics, Massachusetts Institute of Technology, Cambridge, MA 02139, USA}

\author[0000-0002-4265-047X]{Joshua N.\ Winn}
\affiliation{Department of Astrophysical Sciences, Princeton University, Princeton, NJ 08544, USA}




\begin{abstract}

We report the discovery and validation of two long-period giant exoplanets orbiting the early K dwarf TOI-4600 (V=12.6, T=11.9), first detected using observations from the Transiting Exoplanet Survey Satellite (\tess) by the TESS Single Transit Planet Candidate Working Group (TSTPC-WG). The inner planet, TOI-4600 b, has a radius of 6.80$\pm$0.31 $R_{\oplus}$ and an orbital period of 82.69 d. The outer planet, TOI-4600 c, has a radius of 9.42$\pm$0.42 $R_{\oplus}$ and an orbital period of 482.82 d, making it the longest-period confirmed or validated planet discovered by \tess\, to date. We combine \tess\, photometry and ground-based spectroscopy, photometry, and high-resolution imaging to validate the two planets. With equilibrium temperatures of 347 K and 191 K, respectively, TOI-4600 b and c add to the small but growing population of temperate giant exoplanets that bridge the gap between hot/warm Jupiters and the solar system's gas giants. TOI-4600 is a promising target for further transit and precise RV observations to measure masses and orbits for the planets as well as search for additional non-transiting planets. Additionally, with Transit Spectroscopy Metric (TSM) values of $\sim 30$, both planets are amenable for atmospheric characterization with JWST. Altogether will lend insight into the formation and evolution of planet systems with multiple giant exoplanets.

\end{abstract}



\section{Introduction} \label{sec:intro}
There are now over 5000\footnote{\href{https://exoplanetarchive.ipac.caltech.edu/}{NASA Exoplanet Archive}} verified (confirmed and/or validated) exoplanets and they have shed light on the formation, evolution, and general properties of planets. Nonetheless, there are still many open questions regarding these topics and of particular note, are questions regarding long-period planets. As a result of the well-known selection biases \citep{2015ARA&A..53..409W} in the radial velocity and transit methods towards larger, closer-in planets, only $\sim$20\% of the over 5000 verified exoplanets have orbital periods longer than 50 days. While hot Jupiters and, to a lesser extent, warm Jupiters have been well-characterized, the longer-period gas giant regime has been explored much less due to the aforementioned biases in current observing techniques. These planets bridge the gap between the well-studied hot Jupiters and the solar system gas giants, and studying them further and determining their properties will give more insight into planet formation and evolution models \citep{dalba2022}. While radial velocity surveys and \textit{Kepler} have found dozens of long-period giant exoplanets, fewer than 20 have both precisely measured ($>3\sigma$) masses and radii. As a result, not much is known about the planets' compositions or the relationship between the properties of the planets and their host stars.

The NASA \tess\, mission has already led to the discovery of new long-period planets with measured masses and radii \citep{2015JATIS...1a4003R}. Stars in and near \tess's continuous viewing zones are ideal for long-period planet detection because of the long, continuous observing baselines. Planets discovered in these viewing regions include TOI-813 b, TOI-201 b, and TOI-1670 c \citep[][respectively]{2020MNRAS.494..750E, 2021AJ....161..235H, 2022AJ....163..225T}, all of which have orbital periods greater than 40 days. However, for stars located away from the continuous viewing zones, many long-period planets  exhibit only one transit during the timespan of observations. As a result, follow-up radial velocity and photometric observations are needed to determine the orbital period. This has been done to confirm multiple planets, such as TOI-2180 b, NGTS 11 b (TOI-1847 b), and HD 95338 b (TOI-1793 b) \citep[][respectively]{dalba2022, gill2020, diaz2020}. 

Of the aforementioned planets, TOI-1670 c and NGTS 11 b have interior companions of Neptune size or smaller \citep{2022AJ....163..225T, 2022ApJS..259...62I}. Additionally, TOI-201 b may have an interior companion, the super-Earth planet candidate TOI 201.02\footnote{\href{https://tess.mit.edu/toi-releases}{https://tess.mit.edu/toi-releases}}. The presence, or lack thereof, of companions to these planets give insight into the formation and evolution of these systems. There are various formation and evolution scenarios for hot and warm Jupiters, including in-situ formation \citep{2016ApJ...817L..17B}, disk migration \citep{2014prpl.conf..667B}, and high-eccentricity migration \citep{2016ApJ...829..132P}. In-situ formation and disk migration allow for the existence of both inner and nearby outer companions, while high-eccentricity migration precludes the presence of inner companions in most cases \citep{2016ApJ...825...98H}.

Here we present the discovery and validation of the TOI-4600 system, which consists of an early K dwarf with two long-period giant planets originally detected by TESS: TOI-4600 b and c, with periods of 82.7 d and 482.8 d, respectively. We describe the original \tess\, detections and follow-up observations used to characterize and statistically validate the two planets in Section \ref{sec:observations}. In Section \ref{sec:analysis}, we present a detailed analysis of the data in order to determine the stellar and planetary parameters and detail the statistical validation procedure used to rule out false positive scenarios. We discuss the broader context of this system in the growing population of temperate and cool giant exoplanets and prospects for future observations in Section \ref{sec:discussion}.

\section{Observations} \label{sec:observations}

\subsection{TESS Photometry\label{subsec:tess}}

TOI-4600 (TIC 232608943; V = 12.6) was observed by \tess\, Cameras 3 and 4 for 20 sectors to date. It was observed in only the Full Frame Image (FFIs) observations from Sector 14 to Sector 19 (UT 2019 July 18 to UT 2019 December 24) and Sector 21 to Sector 26 (UT 2020 January 21 to UT 2020 July 4). It was then observed in 2-minute cadence in Sectors 40 and 41  (UT 2021 June 24 to UT 2021 August 20), Sectors 47-49 (UT 2021 December 30 to UT 2022 March 26), and Sectors 51-53 (UT 2022 April 22 to UT 2022 July 9). \tess\, observations were interrupted between each of the 13.7-day long orbits of the satellite when data were downloaded to Earth. The TESS Science Processing Operations Center Pipeline \citep[SPOC;][]{2016SPIE.9913E..3EJ} at NASA Ames Research Center calibrated the FFIs and processed the 2-minute data, producing two light curves per sector called Simple Aperture Photometry (SAP) and Presearch Data Conditioning Simple Aperture Photometry \citep[PDCSAP;][]{2012PASP..124.1000S, 2012PASP..124..985S, 2014PASP..126..100S}, the latter of which is corrected for instrumental signatures, screened for outliers, and corrected for crowding effects. The TESS-SPOC pipeline \citep{2020RNAAS...4..201C} extracted photometry from the SPOC-calibrated FFIs.  

TOI-4600 b was first identified as a planet candidate over a year prior to becoming a \tess\, Object of Interest \citep[TOI 4600.01;][]{2021ApJS..254...39G}, by the TESS Single Transit Planet Candidate Working Group (TSTPC WG). The TSTPC WG focuses on searching light curves produced by the Quick-Look Pipeline \citep[QLP;][]{2020RNAAS...4..204H} for single transit events, and validating and/or confirming those that are true planets, with the aim of increasing the yield of intermediate-to-long-period planets found by \tess\, \citep{2019AJ....157...84V, 2020RNAAS...4..204H}. 

\begin{figure*}[!h]
    \centering
    \includegraphics[width=\textwidth]{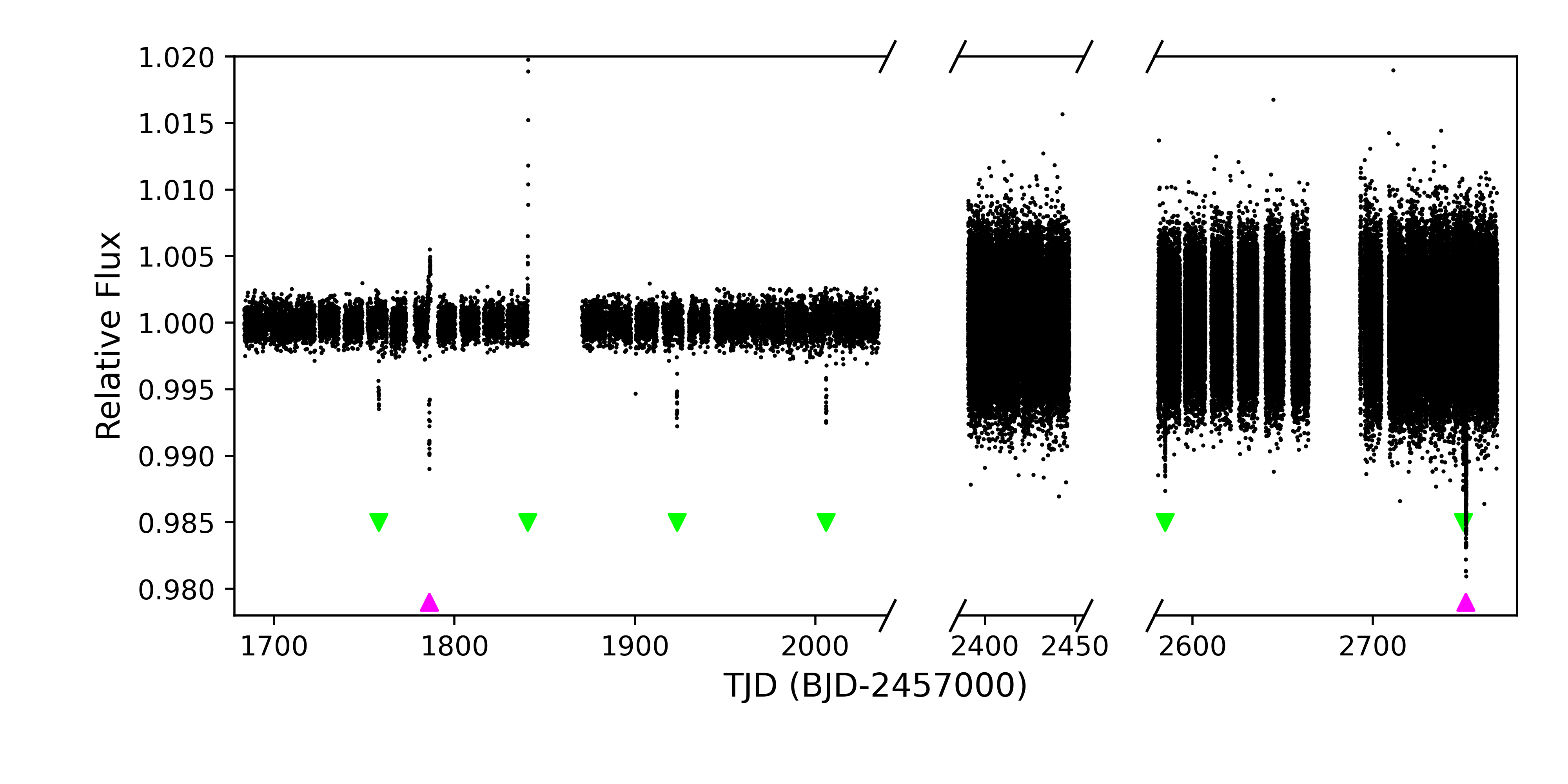}
    \includegraphics[width=\textwidth]{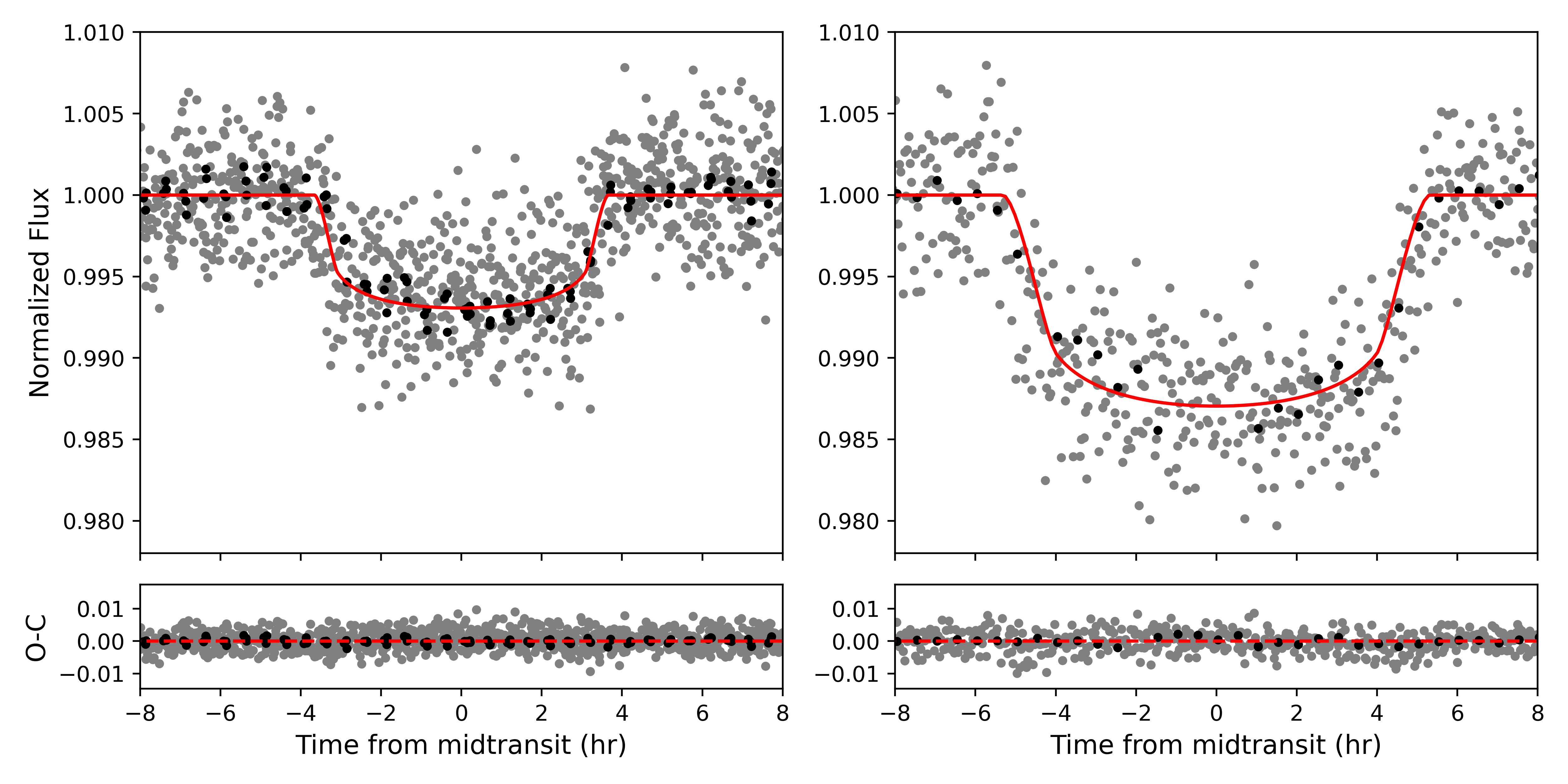}
    \caption{{\it Top:} Full \tess\, PDCSAP light curve of TOI-4600 showing four clear transits of TOI-4600 b and two transits of TOI-4600 c. An additional transit of TOI-4600 b is obscured by a sudden systematic increase in flux due to scattered light near TBJD 1850 while another transit at TBJD 2750 is obscured by a transit of TOI-4600 c that occurs 1.5 days later. Another transit of TOI-4600 b near TBJD 2419 occurred during a downlink gap and thus was not observed by \tess. The apparent difference in the transit depths is due to the different time resolutions, with the left portion showing 30-minute data and the middle and right portions showing 2-minute data. {\it Bottom left:} Phase-folded detrended 2- (gray) and 30-minute (black) \tess\, data and best-fit model for TOI-4600 b. {\it Bottom right:} Same as bottom left but for TOI-4600 c.}
    \label{fig:full_lc}
\end{figure*}

The TSTPC WG identified two $\sim$0.5\% deep transits, one in Sector 16 and one in Sector 22 for TOI-4600 b. Further inspection of the QLP light curve revealed additional transits in Sectors 19 and 25, confirming the period to be 82.69 days (Fig. \ref{fig:full_lc}). It also revealed an addition $\sim$1\% deep transit in Sector 17, which we designate TOI-4600 c. 2-minute cadence data from the Extended Mission have revealed additional transits for both planets, including a transit in Sector 53 of nearly identical depth and duration as that of the transit of TOI-4600 c in Sector 17. We consider this transit in sector 53 to also be of TOI-4600 c, allowing us to constrain the orbital period to only two possible values, as detailed in Section \ref{subsec:monotools}. The \tess\, apertures for TOI-4600 include two neighbors with $\Delta G < 8$ (Fig. \ref{fig:tpfplot}), although both are 6 magnitudes fainter than TOI-4600 and thus too faint to be the sources of the transits. We use \texttt{vetting} \citep{2021RNAAS...5..262H} to rule out centroid offsets for the transits of both TOI-4600 b and c, as detailed in Section \ref{subsec:validation}.

\subsection{Ground-based Photometry\label{subsec:ground}}

\begin{figure}[!h]
    \centering
    \includegraphics[width=\linewidth]{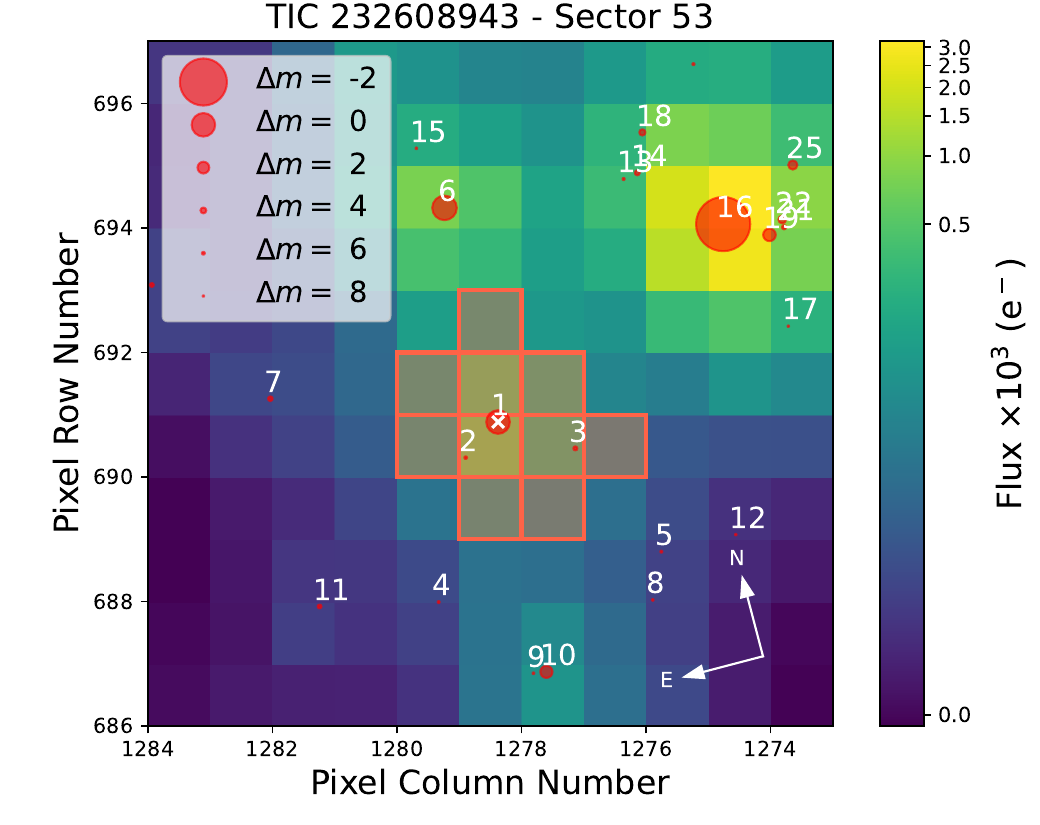}
    \caption{\tess\, target pixel file image of TOI-4600 for Sector 53 made using \texttt{tpfplotter} \citep{2020A&A...635A.128A}.}
    \label{fig:tpfplot}
\end{figure}

The \tess\, pixel scale is $\sim 21\arcsec$ pixel$^{-1}$, and photometric apertures typically extend out to roughly 1 arcminute, which generally results in multiple stars blending in the \tess\, aperture (Figure \ref{fig:tpfplot}). We conducted ground-based photometric follow-up observations of TOI-4600 as part of the {\tt TESS} Follow-up Observing Program\footnote{\href{https://tess.mit.edu/followup}{https://tess.mit.edu/followup}} Sub Group 1 \citep[TFOP;][]{collins:2019} to attempt to rule out or identify nearby eclipsing binaries (NEBs) as potential sources of the \tess\, detection, measure the transit-like event on target to confirm the depth and thus the \tess\, photometric deblending factor, and refine the \tess\, ephemeris. We used the {\tt TESS Transit Finder}, which is a customized version of the {\tt Tapir} software package \citep{Jensen:2013}, to schedule our transit observations. 

\subsubsection{Las Cumbres Observatory\label{subsubsec:lco}}

We observed a transit egress of TOI-4600 b from the Las Cumbres Observatory Global Telescope \citep[LCOGT;][]{Brown:2013} 1.0\,m network node at Teide observatory on the island of Tenerife on UTC 2021 July 23 in Sloan $r'$ band, and a transit ingress (x2)  at McDonald Observatory on UTC 2022 September 10 in the Sloan~$i'$ band with an exposure time of 31s. 

The 1\,m telescopes are equipped with $4096\times4096$ SINISTRO cameras having an image scale of $0\farcs389$ per pixel, resulting in a $26\arcmin\times26\arcmin$ field of view. The images were calibrated by the standard LCOGT {\tt BANZAI} pipeline \citep{McCully:2018}. The differential photometric data were extracted using {\tt AstroImageJ} \citep{Collins:2017} with circular photometric apertures having radius $3\farcs9$. The target star aperture excludes the flux of the nearest Gaia DR3 and TESS Input Catalog neighbor (TIC 232608942) $16\arcsec$ southeast of the target. Both of the partial transits were detected on-target using 3.1\arcsec-3.9\arcsec apertures uncontaminated by known Gaia DR3 and TIC neighbors.

\subsubsection{Wendelstein Observatory\label{subsubsec:wendelstein}}

We observed the same event in Sloan $i'$ band using the Wendelstein Wide-Field-Imager (WWFI), which is a wide field camera consisting of four 4096 $\times$ 4096 CCD detectors and is located on the 2.1\,m Fraunhofer Telescope at the Wendelstein Observatory located near Bayrischzell, Germany \citep{2014ExA....38..213K}. The differential photometric data were extracted using {\tt AstroImageJ}, and a transit egress is detected on-target.

\subsubsection{Kotizarovci Observatory\label{kotizarovci}}

We also observed this transit egress from the Kotizarovci Private Observatory 0.3\,m telescope near Viskovo, Croatia on UTC 2021 July 23 in in Baader R 610\,nm longpass band. The Kotizarovci telescope is equipped with a $765\times510$ pixel SBIG ST7XME camera having an image scale of $1\farcs2$ per pixel, resulting in a $15\arcmin\times10\arcmin$ field of view. The images were calibrated and the differential photometric data were extracted using {\tt AstroImageJ} with circular photometric apertures having radius $7\farcs2$. The target star aperture excludes the flux of the nearest Gaia EDR3 and TESS Input Catalog neighbor (TIC 232608942) $16\arcsec$ southeast of the target. The transit egress was detected on-target.

\subsubsection{Whitin Observatory\label{whitin}}
We observed TOI-4600 on UT 2021 Jul 24 in Sloan $z$ band with the 0.7m telescope at the Wellesley College Whitin Observatory in Massachusetts. We used a $2048\times2048$ FLI Proline camera with a pixel scale of $0\farcs68$. We reduced and extracted time-series differential photometry in an uncontaminated $5\farcs2$ aperture using {\tt AstroImageJ}. The observations were intended to measure an egress of TOI-4600 b on one possible ephemeris; however the ephemeris was later revised and refined with subsequent 2-minute data with the result that we captured only post-egress baseline for the event. 

\subsection{TRES Spectroscopy\label{subsec:spectroscopy}}

We obtained nine reconnaissance spectra of TOI-4600 using the Tillinghast Reflector Echelle Spectrograph (TRES) on the 1.5\,m Tillinghast telescope at the Fred L. Whipple Observatory (FLWO) on Mt. Hopkins in Arizona \citep{gaborthesis}. The nine spectra span dates from 2021 May 31 to 2022 September 28, spanning the full orbit of TOI-4600 b and from phase 0 to 0.5 for TOI-4600 c, and range in signal-to-noise ratios from 21.9 to 35.4. TRES spectra were extracted using procedures outlined in \citet{2010PhDT.......450B} and a multi-order relative velocity analysis was then performed by cross-correlating, order-by-order, the strongest observed spectrum as a template, against all other spectra. We use the Stellar Parameter Classification (SPC) tool to derive spectroscopic parameters ($T_{\rm eff}$, $log\, g$, $v\, sin\, i$, $[m/H]$) for each spectra \citep{2012Natur.486..375B}. We use the spectroscopic parameters to determine physical parameters of the host star, as described in Section \ref{subsec:starparams}. We also use the radial velocities to rule out brown dwarf and stellar mass companions as the sources of the transits, as detailed in Section \ref{subsec:validation}.

\begin{deluxetable}{lccc}
\tablewidth{0pc}
\tabletypesize{\scriptsize}
\tablecaption{
    TRES Relative Radial Velocities
    \label{tab:rvs}
}
\tablehead{
    \multicolumn{1}{c}{Time} &
    \multicolumn{1}{c}{RV$^{a}$}    &
    \multicolumn{1}{c}{Uncertainty} &
    \multicolumn{1}{c}{Instrument}\\
    \multicolumn{1}{c}{BJD} &
    \multicolumn{1}{c}{\ms} &
    \multicolumn{1}{c}{\ms} &
    \multicolumn{1}{c}{}
    }
\startdata
2459365.8122 & 59.8 & 27.3 & TRES \\
2459401.8084 & 131.3 & 30.2 & TRES\\
2459416.7340 & -97.5 & 36.7 & TRES\\
2459468.6876 & 176.2 & 33.3 & TRES\\
2459486.6443 & 177.4 & 24.3 & TRES\\
2459767.7109 & 108.5 & 29.3 & TRES\\
2459771.8213 & 0.0 & 30.2 & TRES\\
2459819.7504 & 133.2 & 31.0 & TRES\\
2459850.7074 & 78.1 & 23.6 & TRES 
\enddata
\tablecomments{(a) The Gaia DR3 RV is 10.62 $\pm$ 0.80 \kms\, \citep{2022arXiv220800211G}.}
\label{rv_data}
\end{deluxetable}

\subsection{High Angular Resolution Imaging\label{highres}}
   As part of our standard process for validating transiting exoplanets to assess the possible contamination of bound or unbound companions on the derived planetary radii \citep{ciardi2015}, we observed TOI-4600 with a combination of high-resolution resources including near-infrared adaptive optics (AO) imaging at Palomar Observatory. While the optical observations tend to provide higher resolution, the NIR AO tend to provide better sensitivity, especially to lower-mass stars. The combination of the observations in multiple filters enables better characterization for any companions that may be detected. 

    
    The Palomar Observatory observations of TOI-4600 were made with the PHARO instrument \citep{hayward2001} behind the natural guide star AO system P3K \citep{dekany2013} on 2021-Jun-19 and 2021~Jun~21 in a standard 5-point quincunx dither pattern with steps of 5\arcsec\ in the narrow-band $Br-\gamma$ filter $(\lambda_o = 2.1686; \Delta\lambda = 0.0326~\mu$m).  Each dither position was observed three times, offset in position from each other by 0.5\arcsec\ for a total of 15 frames; with an integration time of 29.7 seconds per frame, respectively for total on-source times of 445.5 seconds. PHARO has a pixel scale of $0.025\arcsec$ per pixel for a total field of view of $\sim25\arcsec$.
    
    The AO data were processed and analyzed with a custom set of IDL tools.  The science frames were flat-fielded and sky-subtracted.  The flat fields were generated from a median average of dark subtracted flats taken on-sky.  The flats were normalized such that the median value of the flats is unity.  The sky frames were generated from the median average of the 15 dithered science frames; each science image was then sky-subtracted and flat-fielded.  The reduced science frames were combined into a single combined image using an intra-pixel interpolation that conserves flux, shifts the individual dithered frames by the appropriate fractional pixels, and median-coadds the frames.  The final resolutions of the combined dithers were determined from the full-width half-maximum of the point spread functions: 0.12\arcsec\ on 2022-Jun-19 and 0.10\arcsec\ on 2022-Jun-21.  
	
\begin{figure*}
    \centering
    \includegraphics[width=0.49\textwidth]{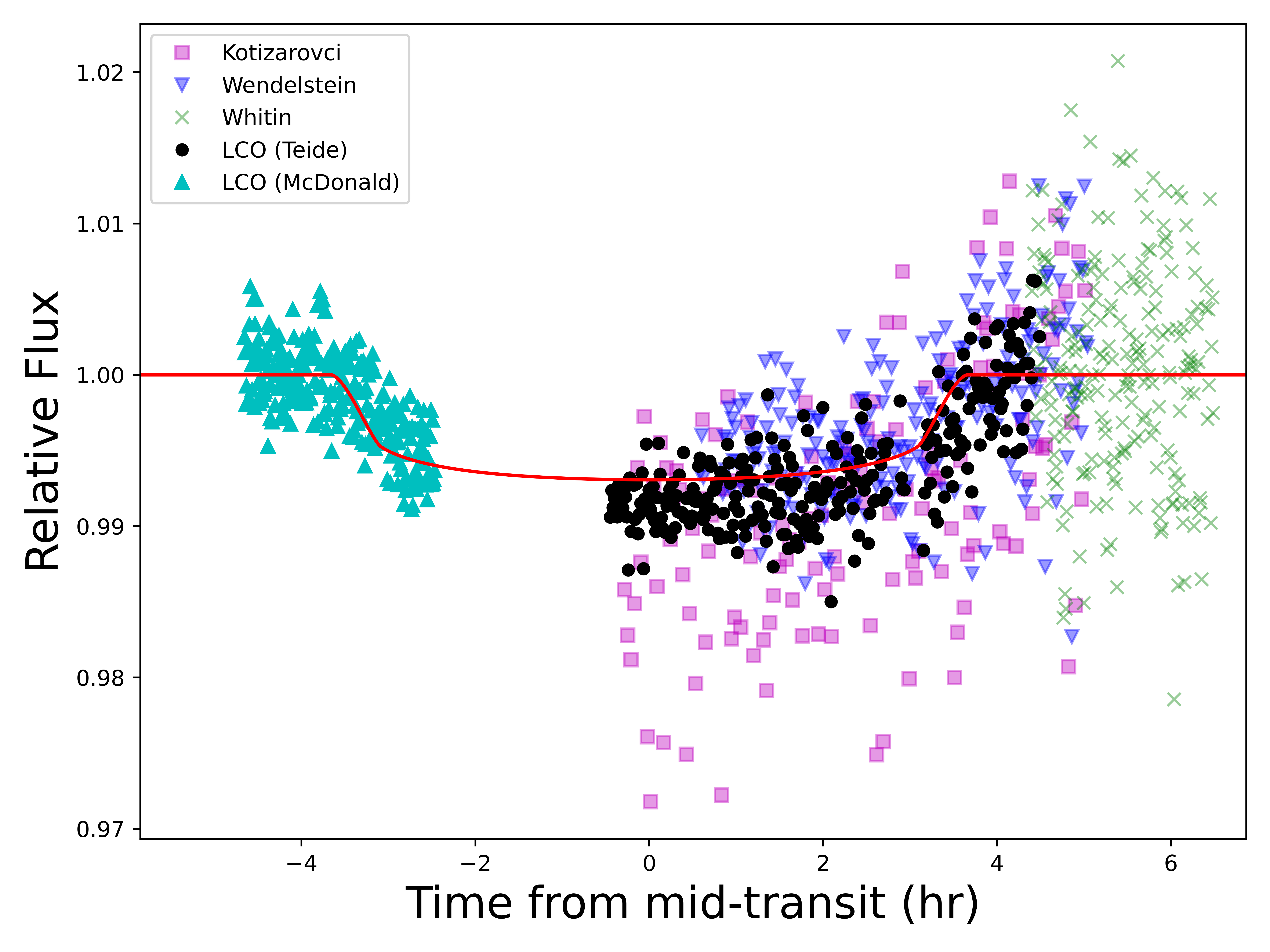}
    \includegraphics[width=0.49\textwidth]{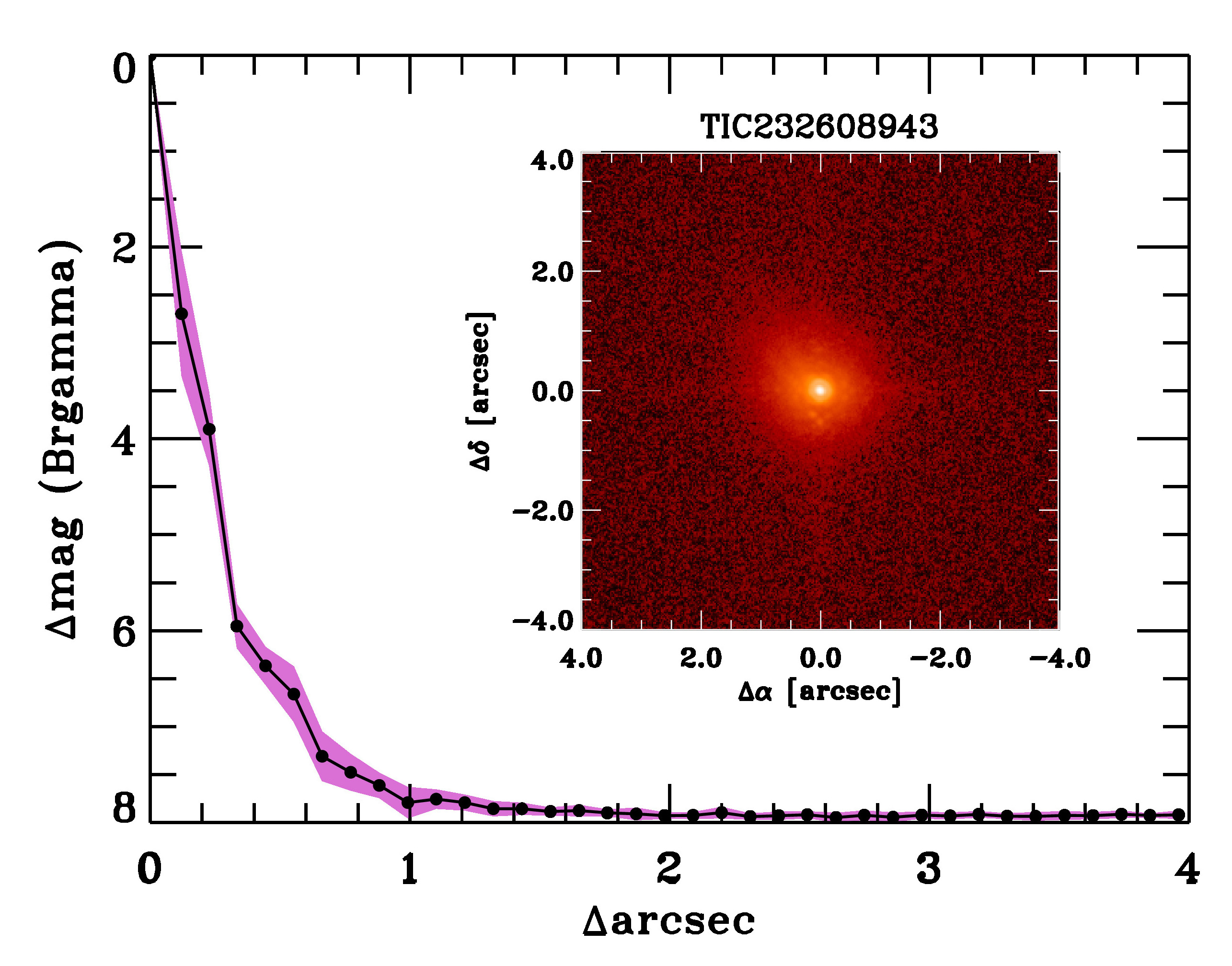}
    \includegraphics[width=\textwidth]{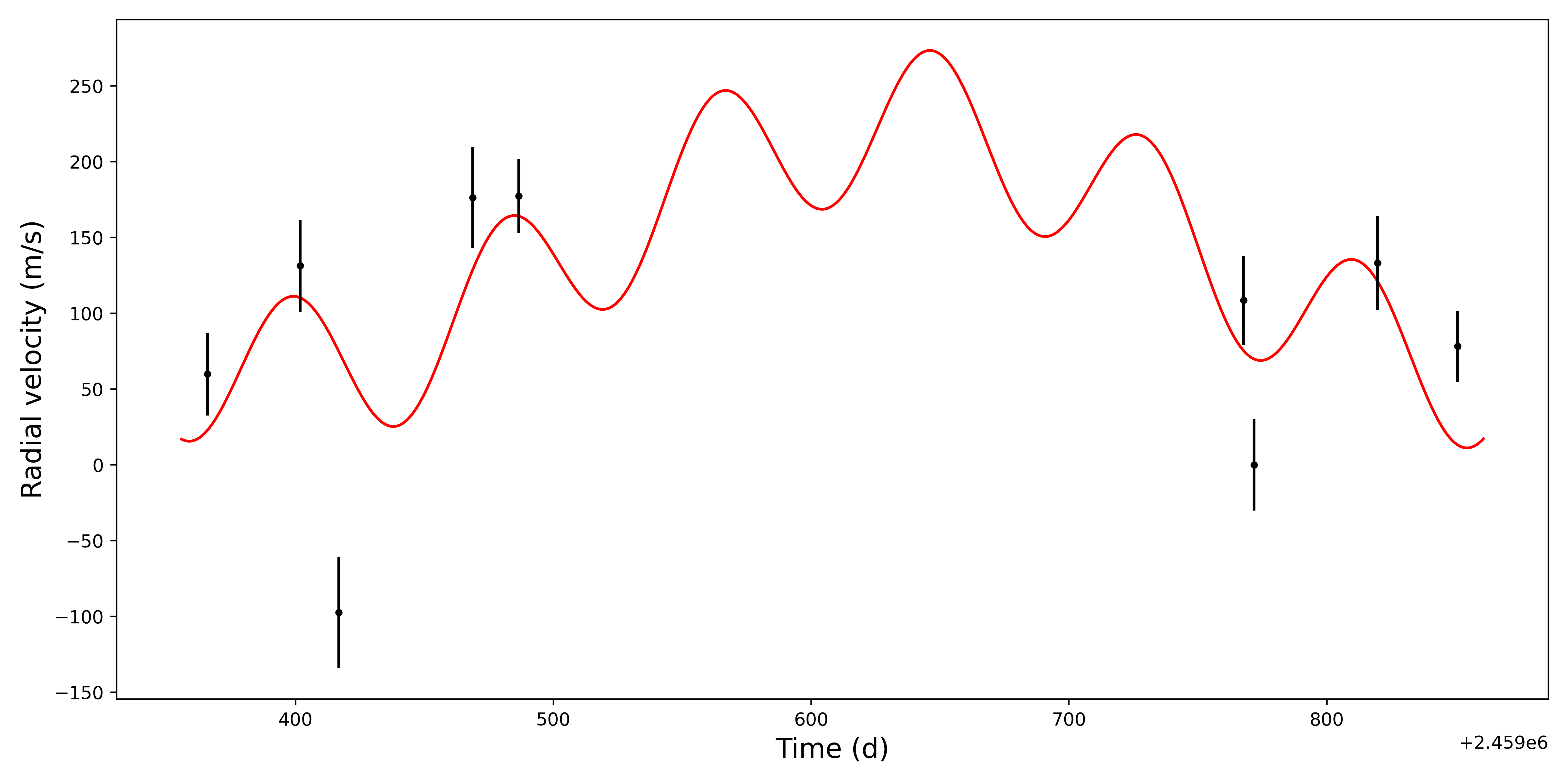}
    \caption{{\it Top left:} Ground based photometry from LCO, Wendelstein Observatory, Kotizarovci Observatory, and Whitin Observatory showing a transit from TOI-4600 b with the best-fit model overlaid. {\it Top right:} Palomar NIR AO imaging and sensitivity curves for TOI-4600 taken in the Br$\gamma$ filter. The images were taken in good seeing conditions, and we reach a contrast of $\sim 7$ magnitudes fainter than the host star within 0.\arcsec5. {\it Inset:} Image of the central portion of the data, centered on the star. {\it Bottom:} Radial velocities extracted from TRES reconnaissance spectra taken over the span of over a year and best-fit \texttt{radvel} model.}
    \label{fig:followup}
\end{figure*}

	The sensitivities of the final combined AO image were determined by injecting simulated sources azimuthally around the primary target every $20^\circ $ at separations of integer multiples of the central source's FWHM \citep{furlan2017}. The brightness of each injected source was scaled until standard aperture photometry detected it with $5-\sigma $ significance. The resulting brightness of the injected sources relative to TOI-4600 set the contrast limits at that injection location. The final $5-\sigma $ limit at each separation was determined from the average of all of the determined limits at that separation and the uncertainty on the limit was set by the rms dispersion of the azimuthal slices at a given radial distance.  The final sensitivity curve for the Palomar data is shown in Figure~\ref{fig:followup}. No additional stellar companions were detected on either night of observing.
    
    \subsection{Gaia Assessment}
    In addition to the high resolution imaging, we have utilized Gaia to identify any wide stellar companions that may be bound members of the system. Typically, these stars are already in the TESS Input Catalog and their flux dilution to the transit has already been accounted for in the transit fits and associated derived parameters. Based upon similar parallaxes and proper motions \citep{mugrauer2020,mugrauer2021}, there are no additional widely separated companions identified by Gaia.
    
    Additionally, the Gaia DR3 astrometry provides additional information on the possibility of inner companions that may have gone undetected by either Gaia or the high resolution imaging. The Gaia Renormalised Unit Weight Error (RUWE) is a metric, similar to a reduced chi-square, where values that are $\lesssim 1.4$  indicate that the Gaia astrometric solution is consistent with the star being single whereas RUWE values $\gtrsim 1.4$ may indicate an astrometric excess noise, possibly caused the presence of an unseen companion \citep[e.g., ][]{ziegler2020}.  TOI-4600 has a Gaia DR3 RUWE value of $\sim 1$ indicating that the astrometric fits are consistent with the single star model.   

\section{Data Analysis} \label{sec:analysis}

\subsection{Host Star Parameters\label{subsec:starparams}}

\subsubsection{MESA Isochrones and Stellar Tracks Analysis}

As noted in section \ref{subsec:spectroscopy}, we use the SPC tool on the $TRES$ spectra to derive initial values for the spectral parameters of the host star, obtaining a $T_{\rm eff}$ = 5200 K, log $g$ = 4.6, and [m/H] = 0.15. We then used the spectroscopic parameters along with the Gaia DR3 parallax and magnitudes ($G$, $B_P$, $R_P$), 2MASS magnitudes ($J$, $H$, $K_S$), and WISE magnitudes ($W_1$, $W_2$, $W_3$) to perform an isochrone fit in order to further constrain the spectroscopic parameters and derive physical parameters for the host star. The spectroscopic parameters, parallax, and magnitudes are used as priors to determine the goodness of fit. We use the \texttt{isochrone} package \citep{2015ascl.soft03010M} to generate the isochrone models used to sample the stellar parameters and find the best-fit parameters by using a Markov Chain Monte Carlo (MCMC) routine using the \textbf{emcee} package \citep{2013PASP..125..306F}. The routine consists of 40 independent walkers each taking 25000 steps, of which the first 2000 are discarded as burn-in. \texttt{emcee} checks whether the chains have converged by determining if the total number of steps is at least 100 times the autocorrelation time, as is the case with the routine here. The fitted spectroscopic parameters and derived physical parameters, including stellar age, for the host star are reported in Table \ref{system_info}, and are consistent with a K1V star \citep{2013ApJS..208....9P}. We inflate the uncertainties of the parameters derived from the isochrone modeling in order to correct for the underestimated uncertainties commonly seen with stellar evolutionary models, adding the systematic uncertainty floors from \citep{2022ApJ...927...31T} in quadrature to the uncertainties from the isochrone modeling.

\subsubsection{Spectral Energy Distribution Analysis}

\begin{figure}[!h]
    \centering
    \includegraphics[width=\linewidth,trim=400 400 400 400,clip]{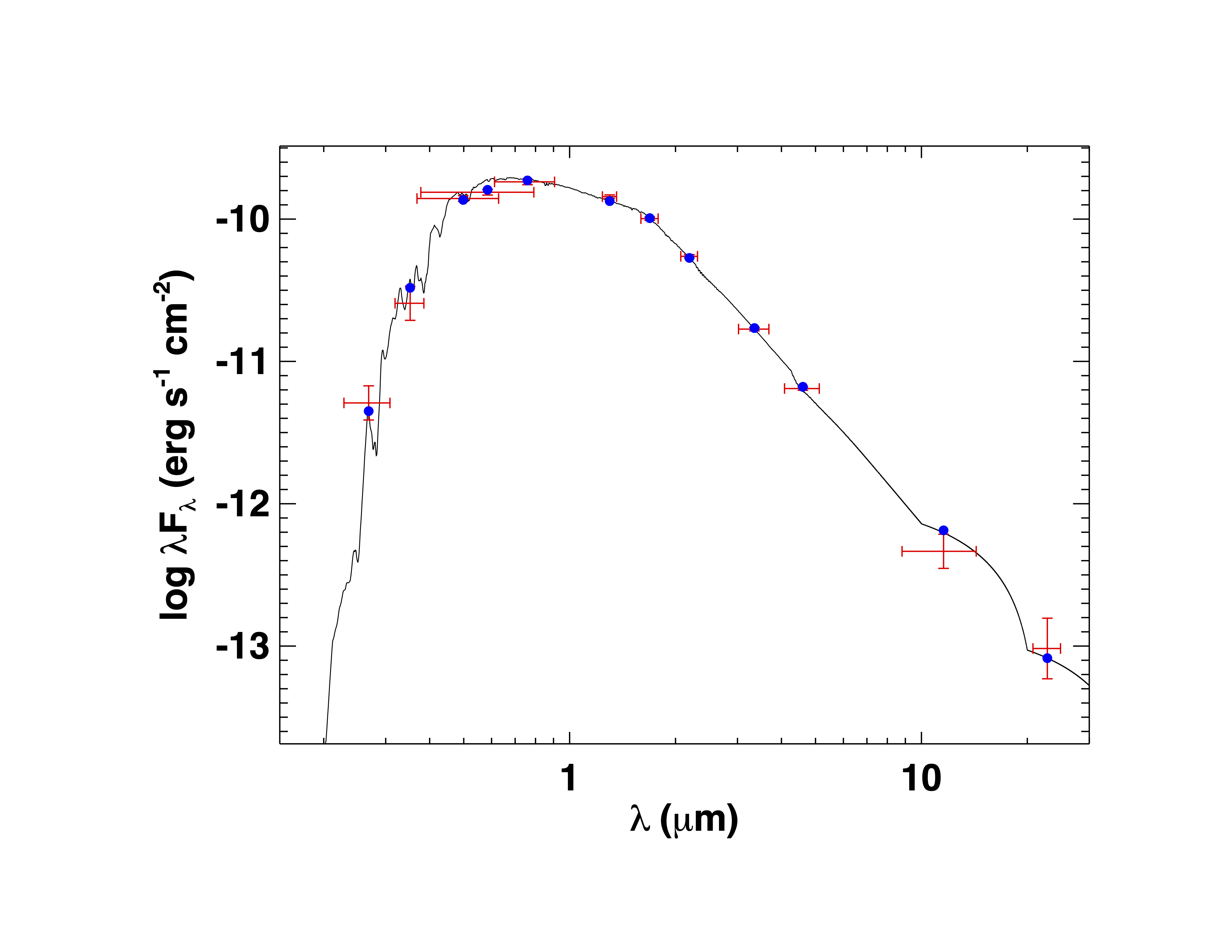}
    \caption{Spectral energy distribution of TOI-4600. Red symbols represent the observed photometric measurements, where the horizontal bars represent the effective width of the passband. Blue symbols are the model fluxes from the best-fit Kurucz atmosphere model (black).}
    \label{fig:sed}
\end{figure}

As an independent determination of the basic stellar parameters, we performed an analysis of the broadband spectral energy distribution (SED) of the star together with the {\it Gaia\/} DR3 parallax \citep[with no systematic offset applied; see, e.g.,][]{StassunTorres:2021}, in order to determine an empirical measurement of the stellar radius, following the procedures described in \citet{Stassun:2016,Stassun:2017,Stassun:2018}. We pulled the $JHK_S$ magnitudes from {\it 2MASS}, the W1--W4 magnitudes from {\it WISE}, and the $G, G_{\rm BP}, G_{\rm RP}$ magnitudes from {\it Gaia}. We also used the UVW1 and $U$-band measurements from the {\it Swift\/} satellite. Together, the available photometry spans the full stellar SED over the wavelength range 0.3--20~$\mu$m (see Figure~\ref{fig:sed}).  

We performed a fit using Kurucz stellar atmosphere models, with the free parameters being the effective temperature ($T_{\rm eff}$), surface gravity ($\log g$), and metallicity ([Fe/H]). The remaining free parameter is the extinction $A_V$, which we limited to maximum line-of-sight value from the Galactic dust maps of \citet{Schlegel:1998}. The resulting fit (Figure~\ref{fig:sed}) has a reduced $\chi^2$ of 1.3 with best-fit $A_V = 0.04 \pm 0.04$, $T_{\rm eff} = 5075 \pm 75$~K, $\log g = 4.5 \pm 0.5$, and [Fe/H] = $-0.2 \pm 0.3$. Integrating the model SED gives the bolometric flux at Earth, $F_{\rm bol} = 2.791 \pm 0.066 \times 10^{-10}$ erg~s$^{-1}$~cm$^{-2}$. Taking the $F_{\rm bol}$ and $T_{\rm eff}$ together with the {\it Gaia\/} parallax, gives the stellar radius, $R_\star = 0.827 \pm 0.026$~R$_\odot$. In addition, we can estimate the stellar mass from the empirical relations of \citet{Torres:2010}, giving $M_\star = 0.81 \pm 0.10$~M$_\odot$. These parameters are consistent with those derived using the isochrone models at the 1$-\sigma$ level. 

\def\arraystretch{1.25}
\begin{deluxetable*}{lcc||lcc}
\tablewidth{0pc}
\tabletypesize{\scriptsize}
\tablecaption{
    System Information
    \label{tab:info}
}
\tablehead{
    \multicolumn{1}{c}{Stellar Parameter} &
    \multicolumn{1}{c}{Value} &
    \multicolumn{1}{c}{Source} &
    \multicolumn{1}{c}{Planet Parameter} & 
    \multicolumn{1}{c}{TOI-4600 b} & 
    \multicolumn{1}{c}{TOI-4600 c}
}
\startdata
TIC & 232608943 &  TIC V8$^a$ & \multicolumn{3}{l}{\textit{Fitted parameters}} \\  
R.A. & 17:13:48.09 & Gaia DR3$^b$ & $q_{1, \tess}$ & 0.31$_{-0.14}^{+0.24}$ & - \\ 
Dec. & +64:33:58.22 & Gaia DR3$^b$ & $q_{2, \tess}$ & 0.34$_{-0.21}^{+0.33}$ & - \\
$\mu_{ra}$ (mas yr$^{-1}$)  & 7.625 $\pm$ 0.012 & Gaia DR3$^b$ & $R_p / R_{\star}$ & 0.0769$_{-0.0017}^{+0.0021}$ & 0.1067$_{-0.0026}^{+0.0024}$ \\
$\mu_{dec}$ (mas yr$^{-1}$) & 0.266 $\pm$ 0.016 & Gaia DR3$^b$ & $(R_{\star} + R_p) / a$ & 0.01161$_{-0.00052}^{+0.00060}$ & 0.00362$_{-0.00015}^{+0.00017}$ \\ 
Parallax (mas) & 4.620 $\pm$ 0.011 & Gaia DR3$^b$ & $\cos{i}$ & 0.0041$_{-0.0025}^{+0.0016}$ & 0.00169 $_{-0.00056}^{+0.00038}$ \\ 
Epoch & 2016.0 &  Gaia DR3$^b$ & $T_0$ (BJD - 2,457,000) & 2750.1421$_{-0.0019}^{+0.0020}$ & 2751.6008 $\pm$ 0.0020 \\ 
$B$ (mag)    & 13.487 $\pm$ 0.048 & AAVSO DR9$^c$ & $P$   $\mathrm{(d)}$ & 82.6869$\pm$0.0003 & 482.8191$_{-0.0017}^{+0.0018}$ \\
$V$ (mag)    & 12.577 $\pm$ 0.034 & AAVSO DR9$^c$ & $\sqrt{e}\, cos\, \omega$ & 0.01$_{-0.52}^{+0.55}$ & 0.04$_{-0.55}^{+0.54}$ \\
$Gaia$ (mag) & 12.43401 $\pm$ 0.00027 & Gaia DR3$^b$ & $\sqrt{e}\, sin\, \omega$ & -0.30$_{-0.16}^{+0.20}$ & 0.10$_{-0.22}^{+0.20}$ \\
$B_P$ & 12.92036 $\pm$ 0.00285 & Gaia DR3$^b$ & ln $\sigma_{GP,s}$ & -6.92$_{-0.37}^{+0.66}$ & - \\
$R_P$ & 11.82667 $\pm$ 0.00380 & Gaia DR3$^b$ & ln $\rho_{GP,s}$ & -0.33$_{-0.73}^{+0.82}$ & - \\
\tess\, (mag) & 11.8787 $\pm$ 0.006 & TIC V8$^a$ & offset$_{GP,s}$ & 0.0007$_{-0.0007}^{+0.0005}$ & - \\
$J$ (mag)    & 11.075 $\pm$ 0.023 & 2MASS$^d$ & ln $\sigma_{GP,l}$ & -6.48$_{-0.30}^{+0.41}$ & - \\
$H$ (mag)    & 10.659 $\pm$ 0.021 & 2MASS$^d$ & ln $\rho_{GP,l}$ & 0.59$_{-0.58}^{+0.68}$ & - \\
$K_S$ (mag)  & 10.571 $\pm$ 0.018 & 2MASS$^d$ & offset$_{GP,l}$ & 0.001 $\pm$ 0.001 & - \\
$M_{\star}$ ($M_{\odot}$)  & 0.89 $\pm$ 0.05 & This work & ln $\sigma_{TESS,s}$ & -5.85 $\pm$ 0.02 & - \\
$R_{\star}$ ($R_{\odot}$)   & 0.81 $\pm$ 0.03 & This work & ln $\sigma_{TESS,l}$ & -7.10 $\pm$ 0.05 & - \\
$\rho_{\star}$ (g cm$^{-3}$) & 2.36 $\pm$ 0.29 & This work & & & \\ 
$L_{\star}$ ($L_{\odot}$)   & 0.42 $\pm$ 0.02 & This work & \multicolumn{3}{l}{\textit{Derived parameters}} \\
$T_{eff}$ (K)      & 5170 $\pm$ 120 & This work & $i$ ($^{\circ}$) & 89.76$_{-0.10}^{+0.14}$ & 89.90$_{-0.02}^{+0.03}$ \\ 
$[$Fe/H$]$         & 0.16 $\pm$ 0.08 & This work & $a$ (AU) & 0.349 $\pm$ 0.021 & 1.152 $\pm$ 0.068 \\ 
log $g$          & 4.57 $\pm$ 0.02 & This work & $b$ & 0.39$_{-0.24}^{+0.17}$ & 0.46$_{-0.19}^{+0.10}$ \\
Age (Gyr)       & 2.3$^{+2.8}_{-1.6}$  & This work & $T_\mathrm{tot}^e$ (h) & 7.54$_{-0.13}^{+0.14}$ & 11.14$_{-0.19}^{+0.20}$ \\  
$v\, \text{sin} i$ (\kms) & 2.5 $\pm$ 1.3 & This work & $T_\mathrm{full}^f$  (h) & 6.26$_{-0.20}^{+0.15}$ & 8.47$_{-0.33}^{+0.30}$ \\ 
& & & $R_p$ ($R_{\oplus}$) & 6.80$_{-0.30}^{+0.31}$ & 9.42$_{-0.41}^{+0.42}$ \\
& & & $R_p$ ($R_J$) & 0.607$_{-0.026}^{+0.028}$ & 0.841 $\pm$ 0.037 \\
& & & $e$ & 0.25$_{-0.17}^{+0.33}$ & 0.21$_{-0.14}^{+0.29}$ \\
& & & $\omega$ & 260$_{-57}^{+70}$ & 142$_{-116}^{+148}$ \\
& & & $T_\mathrm{eq}^g$ (K) & 347$_{-11}^{+12}$ & 191 $\pm$ 6 \\
\enddata
\tablecomments{(a) \cite{2018AJ....156..102S}. (b) \cite{2021AA...649A...1G}. (c) \cite{2016yCat.2336....0H}. (d) \cite{2003yCat.2246....0C}. (e) From 1st to last (4th) contacts. (f) From 2nd to 3rd contacts. (g) Assuming an albedo of 0.3 and emissivity of 1.}
\label{system_info}
\end{deluxetable*}

\subsection{Orbital Period of the Outer Candidate\label{subsec:monotools}}

The second transit of TOI-4600 c observed in sector 53 meant the possible periods were reduced to a finite number of aliases of the separation between the two transits. As an initial check that the two transit events are caused by the same planet, we model each event independently using \texttt{allesfitter} \citep[][see Section \ref{subsec:allesfitter_fit} for a more detailed description]{2021ApJS..254...13G} in order to obtain estimates for the planet radius and orbital period, assuming a circular orbit. We obtain a planet radius of $9.42^{+0.37}_{-0.29}\, R_{\oplus}$ and $9.53^{+0.28}_{-0.25}\, R_{\oplus}$ for the first and second transits, respectively. For the orbital periods, we obtain estimates of $360^{+170}_{-100}$ d and $324^{+84}_{-58}$ d, respectively. We find these values agree with each other to within 1-$\sigma$ and therefore we safely assume from this point on that these two events are caused by the same source and hence planetary candidate.

In order to assess the potential orbital periods for TOI-4600 c, we first fitted the available \tess\, transits jointly using the \texttt{MonoTools} method outlined in \citet{2022arXiv220303194O}. This fits impact parameter, transit duration and radius ratio in a way that is agnostic of the exoplanet period using the \texttt{exoplanet} fitting package \citep{exoplanet:joss}. When combined with stellar parameters such as bulk density \citep[in this case from the TICv8 catalog;][]{2019AJ....158..138S}, the transit model allows us to derive an instantaneous transverse planetary velocity and therefore, for each allowed period alias, to compute a marginalised probability distribution. This uses a combination of the model likelihood, and priors from a combination of window function and occurrence rate \citep[$P^{-2}$;][]{2018RNAAS...2..223K}, a geometric transit probability ($p_g \propto a^{-1} \sim P^{-2/3}$), and an eccentricity prior \citep{2019AJ....157...61V} applied using the eccentricity distribution implied by the ratio of the transverse planetary velocity to the circular velocity at that period alias. 
The stability of each possible orbit with respect to other planets with known periods in the system is also considered by suppressing the probabilities of any orbit which passes inside the Hill radii of inner planets.

Through \texttt{MonoTools}, we used the TESS SPOC HLSP FFI lightcurves (S14-S26), the 2-minute SPOC light curves (S40-41, S47-49), and light curves extracted from the 10-minute TICA FFIs (S51-53) to analyze the system. We flattened the data using a basis spline set by a 1.2d knot distance and with identified transits from planets b \& c masked. We then clipped the \tess\, photometry to windows around each transit with 4.5 transit durations, thereby improving the speed of sampling. Sampling was performed using the PyMC3 implementation of Hamiltonian Monte Carlo \citep{2016ascl.soft10016S} using 4 chains and producing 2000 unique samples, with low Rubin-Gelman statistics ($\hat{r}$) ensuring that the chains were well-mixed. The resulting log probabilities are then marginalised using the sum of the probability across all period aliases, and the final log probabilities were calculated by summing the probabilities across all samples (assuming equal weight for each independent model draw) and period aliases. Using \texttt{MonoTools}, we find again that the two transits are related and we find that there are only two periods permitted as a result of the vast data coverage --- 482.82 and 965.64 d --- potential transits at all other periods  are ruled out by TESS observations. \texttt{Monotools} assigns these two possible periods with probabilities of 99.97\% and 0.03\%, respectively, assuming the eccentricity distribution of multi-planet systems from \citet{2019AJ....157...61V}. We note that when assuming a more general eccentricity distribution from \citep{2013MNRAS.434L..51K}, the 482.82 d alias is preferred by a factor of only $\sim6.5$. As a result of this analysis, we hereafter assume an orbital period of 482.82 d for TOI-4600 c. This period can be definitively confirmed with an additional transit observation, with the next predicted transit of only the 482.82 d alias occurring on UT 2023 October 16. Given the previous ground-based observations of the smaller inner planet, the predicted transit of TOI-4600 c should be readily detected.

\subsection{False Positive Scenarios and Statistical Validation\label{subsec:validation}}

We are able to rule out the various false positive scenarios using the original \tess\, photometry and subsequent ground-based photometry, spectroscopy, and imaging. We analyze the in- and out-of-transit photocenters, or centroids, using {\tt vetting} \citep{2021RNAAS...5..262H} to rule out nearby eclipsing binaries (NEBs) as the source of the transits for both planets. {\tt vetting} calculates a p-value for each \tess\, sector to determine whether a transit is on-target or not. For TOI-4600 c, the p-values for both sectors is greater than 0.05, indicating the transits are on-target. For TOI-4600 b, all sectors except sector 19 have p-values greater than 0.05. The transit  in that sector is contaminated by scattered light causing a false offset. We are able to rule out stellar masses for both planets using the TRES spectra. We fit the radial velocities using the \texttt{radvel} package \citep{2018PASP..130d4504F} in order to obtain mass limits for both planets. We assume circular orbits for both planets and fix the period and epoch to the best-fit values listed in Table \ref{tab:info} for both. We fit only the radial velocity semi-amplitudes of both planets, a jitter term for TRES, and relative offset term. We use a MCMC routine consisting of $4\times10^6$ steps with uniform priors for all parameters. We obtain 3-$\sigma$ upper mass limits of 3.02 $M_J$ and 9.27 $M_J$ for TOI-4600 b and c, respectively. 

We use \texttt{triceratops} \citep{2020ascl.soft02004G, 2021AJ....161...24G} to calculate false-positive probabilities (FPP) and nearby false-positive probabilities (NFPP) for each candidate in the system, including the contrast curve from our high-resolution imaging in order to provide additional constraints. False-positive scenarios include an eclipsing binary on target, on a background star, or an unseen companion as well as a transiting planet on a background star or unseen companion. Nearby false-positive scenarios include a transiting planet or eclipsing binary on a nearby star. We calculate FPP values of 0.0107 and 0.0205 for b and c, respectively. Due to the lack of sufficiently bright nearby neighbors, the NFPP values for both of the candidates defaults to 0. While the individual FPPs of both TOI-4600 b and c exceed the nominal maximum value of 0.015 stated by \citet{2021AJ....161...24G} as the threshold to be considered statistically validated, multi-planet system candidates have been found to have much lower false-positive rates than single-planet system candidates. For the \textit{Kepler} mission, the false-positive rate of multi-planet system candidates was $\sim 25$ times lower than that of single-planet system candidates \citep{2012ApJ...750..112L, 2014ApJ...784...44L}. For \tess\, this so-called ``multiplicity boost'' is $\sim 20$ \citep{2021ApJS..254...39G}, meaning the FPP values of both TOI-4600 b and c are well below the previously mentioned 0.015 threshold and the often-used 0.01 threshold \citep{2014ApJ...784...45R, 2015ApJ...809...25M, 2019A&A...627A..66H, 2020MNRAS.499.5416C}. 

The false positive probability values of both planets are dominated by the Secondary Transit Planet (STP) scenario, where the transits originate from a transiting planet orbiting an unresolved bound companion. However, \texttt{triceratops} only uses the contrast curve to determine possible bound companions in its analysis of the STP scenario. In order to incorporate additional data, we use Multi-Observational Limits on Unseen Stellar Companions \citep[MOLUSC;][]{2021AJ....162..128W} in order to generate a sample of potential companions consistent with the combination of the contrast curve, radial velocity data, \textit{Gaia} astrometry (in the form of the RUWE), and \textit{Gaia} imaging. Of the 5 million stars we generated using MOLUSC, only $\sim10\%$ were consistent with the aforementioned data. We then reran \texttt{triceratops} using this sample of plausible bound companions and recalculated the FPPs, finding that the STP scenario is now a negligible component of the FPP. We obtain a FPP of $(3.7\pm6.8) \times 10^{-5}$ for b and $(1.1\pm3.9) \times 10^{-7}$ for c, making both statistically validated planets. 

\subsection{Photometric Fit\label{subsec:allesfitter_fit}}

In order to derive the orbital parameters and radii of the two planets, we performed model fitting using the publicly available \texttt{allesfitter} package \citep{2021ApJS..254...13G}. We used MCMC sampling using the \textbf{emcee} package to explore the parameter space and determine the best-fit values from the medians of the posteriors for the following parameters:

\begin{itemize}
    \itemsep0pt 
    \item quadratic stellar limb-darkening parameters $q_1$ and $q_2$, using the transformation from \citet{2013MNRAS.435.2152K}, with uniform priors from 0 to 1
    \item radius ratio, $R_p/R_{\star}$, where p denotes the individual planets, with uniform prior from 0 to 1,
    \item sum of radii divided by the orbital semi-major axis, $(R_{\star} + R_p) / a$, with uniform prior from 0 to 1,
    \item cosine of the orbital inclination, $ \cos{i}$, with uniform prior from 0 to 1, 
    \item orbital period, $P$, with uniform prior centered on the estimate of 82.69 d from \texttt{MonoTools} with a 1 d range,
    \item transit epoch, $T_{0}$, with uniform prior centered on the estimate of 482.82 d from \texttt{MonoTools} with a 1 d range,
    \item eccentricity parameters $\sqrt{e}\, cos\, \omega$ and $\sqrt{e}\, sin\, \omega$, each with uniform prior from -1 to 1, where $e$ is the orbital eccentricity and $\omega$ the argument of periastron,
    \item the hyperparameters $\sigma_{GP}$ and $\rho_{GP}$ and offset$_{GP}$ for a Mat\'{e}rn 3/2 kernel used to model the red noise for the 2- and 30-minute \tess\, data individually (denoted $s$ and $l$, respectively)
    \item white noise scaling terms for the 2- and 30-minute \tess\, data, $\sigma_{TESS, s}$ and $\sigma_{TESS, l}$.

\end{itemize}

We initialized the MCMC with 200 walkers, performing 2 preliminary runs of 1000 steps per walker to obtain higher-likelihood initial guesses for the nominal run of 40000 steps per walker. We then discarded the first 10000 steps for each chain as burn-in phase before thinning the chains by a factor of 100 and calculating the final posterior distributions. The values and uncertainties of the fitted and derived parameters listed in Table \ref{system_info} are defined as the median values and 68\% confidence intervals of the posterior distributions, respectively. The best-fit transit model light curves for the planets are shown in Figure \ref{fig:full_lc}.

\section{Discussion} \label{sec:discussion}

We derive radii of $6.80_{-0.30}^{+0.31}$ and $9.42_{-0.41}^{+0.42}$ $R_{\oplus}$ for TOI-4600 b and c, respectively, from a fit of the \tess\, photometry. Combined with respective equilibrium temperatures of approximately 350 K and 190 K, this makes TOI-4600 b a temperate sub-Saturn and TOI-4600 c a cold Saturn. While the eccentricities of both planets are consistent with 0, the transit duration of planet c may hint at a non-zero eccentricity. The analysis of the individual transits of planet c described in Section \ref{subsec:monotools} predicted an orbital period in the range of 320-360 days when assuming a circular orbit, compared to the true orbital period of 482 d. This disparity between the predicted and observed values is due to the transit duration being shorter than expected for a circular orbit at 482 d. However, due to the degeneracies between the impact parameter, orbital inclination, and transit duration, it is difficult to place a statistically significant constraint on the planet's eccentricity with photometry alone. 

\begin{figure*}[!h]
\begin{minipage}[c]{.515\textwidth}
  \vspace*{\fill}
  \centering
  \includegraphics[width=\textwidth]{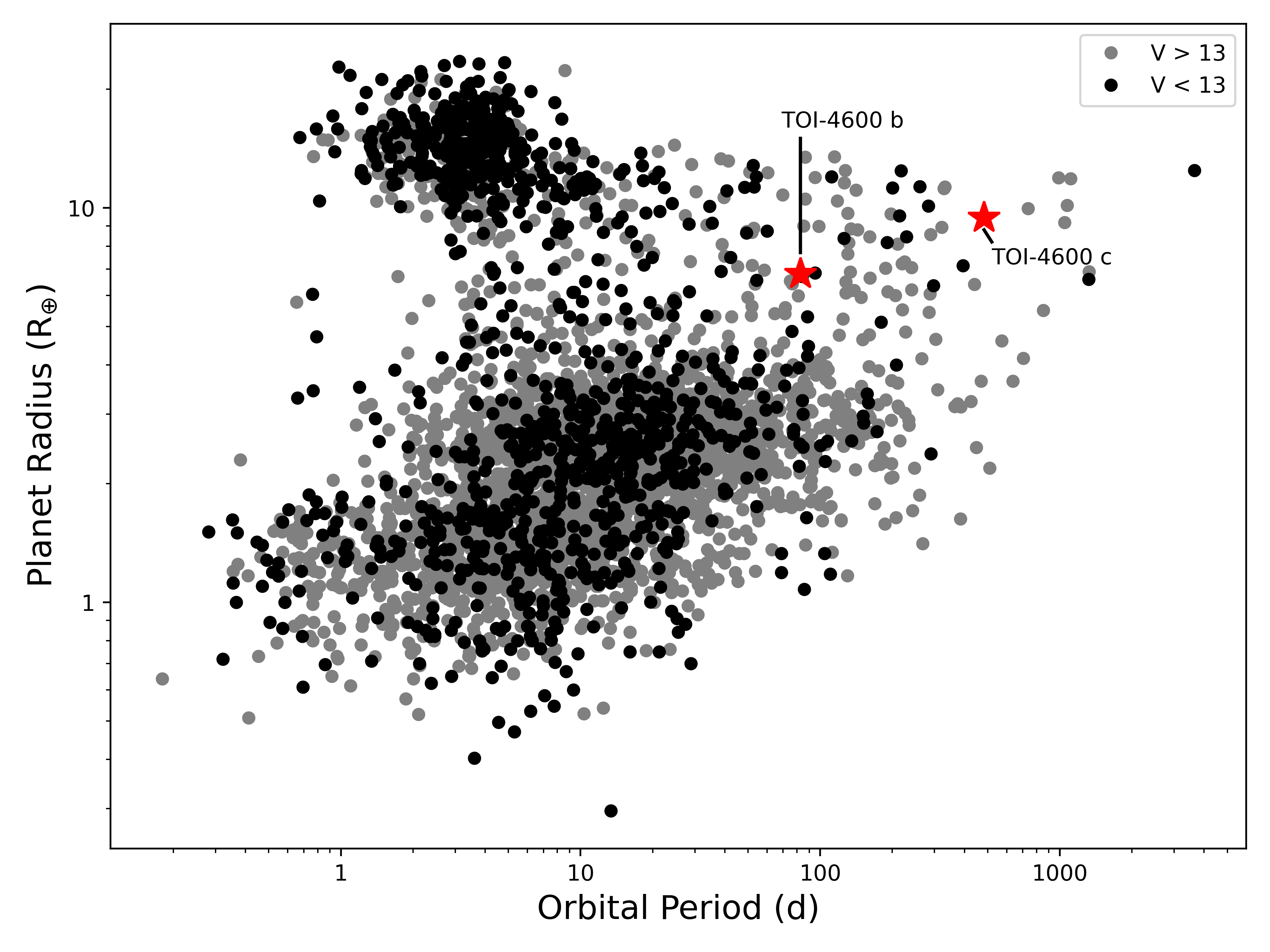}
  \includegraphics[width=\textwidth]{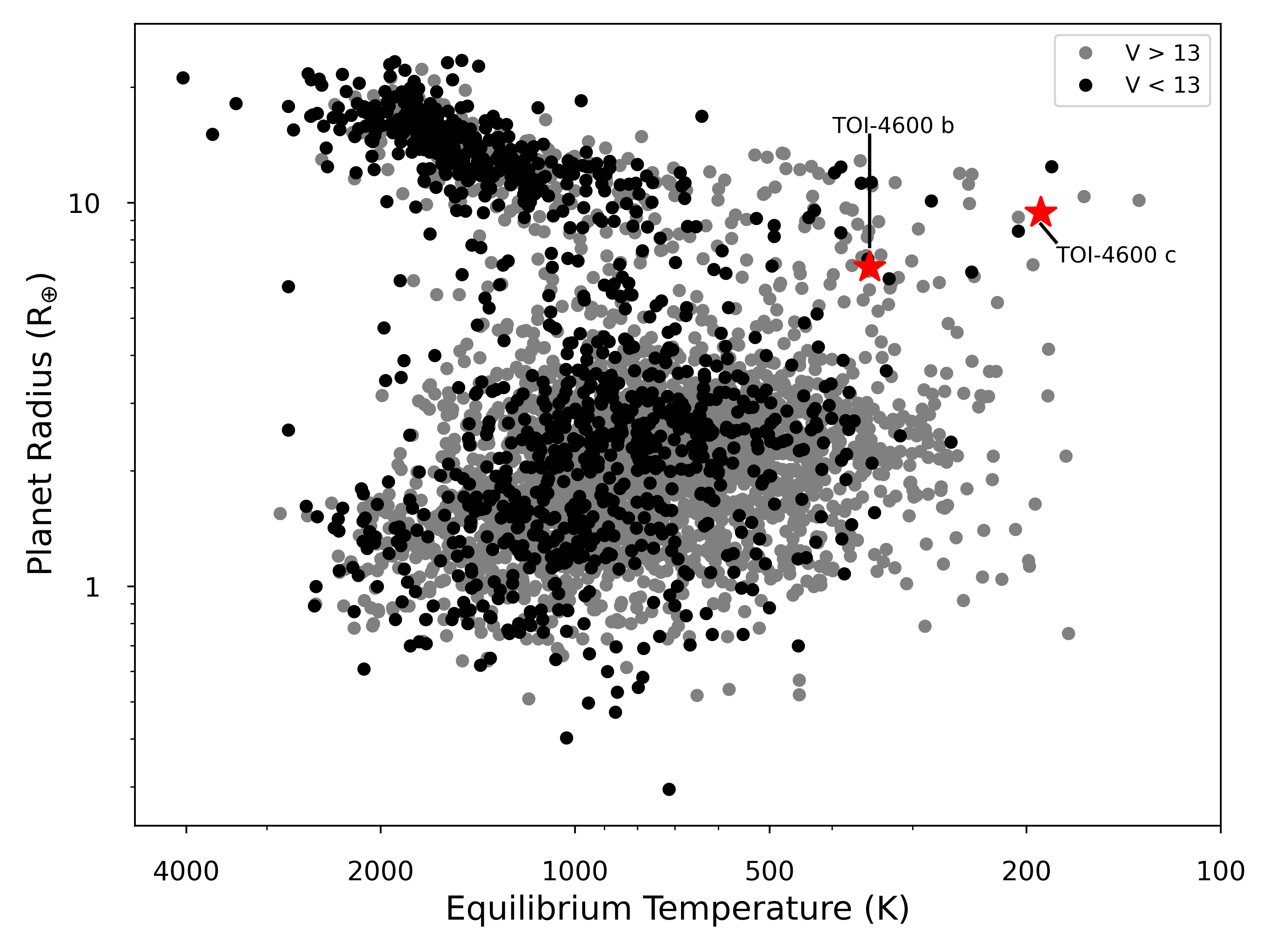}
\end{minipage}%
\begin{minipage}[c]{.485\textwidth}
  \vspace*{\fill}
  \centering
  \includegraphics[width=\textwidth]{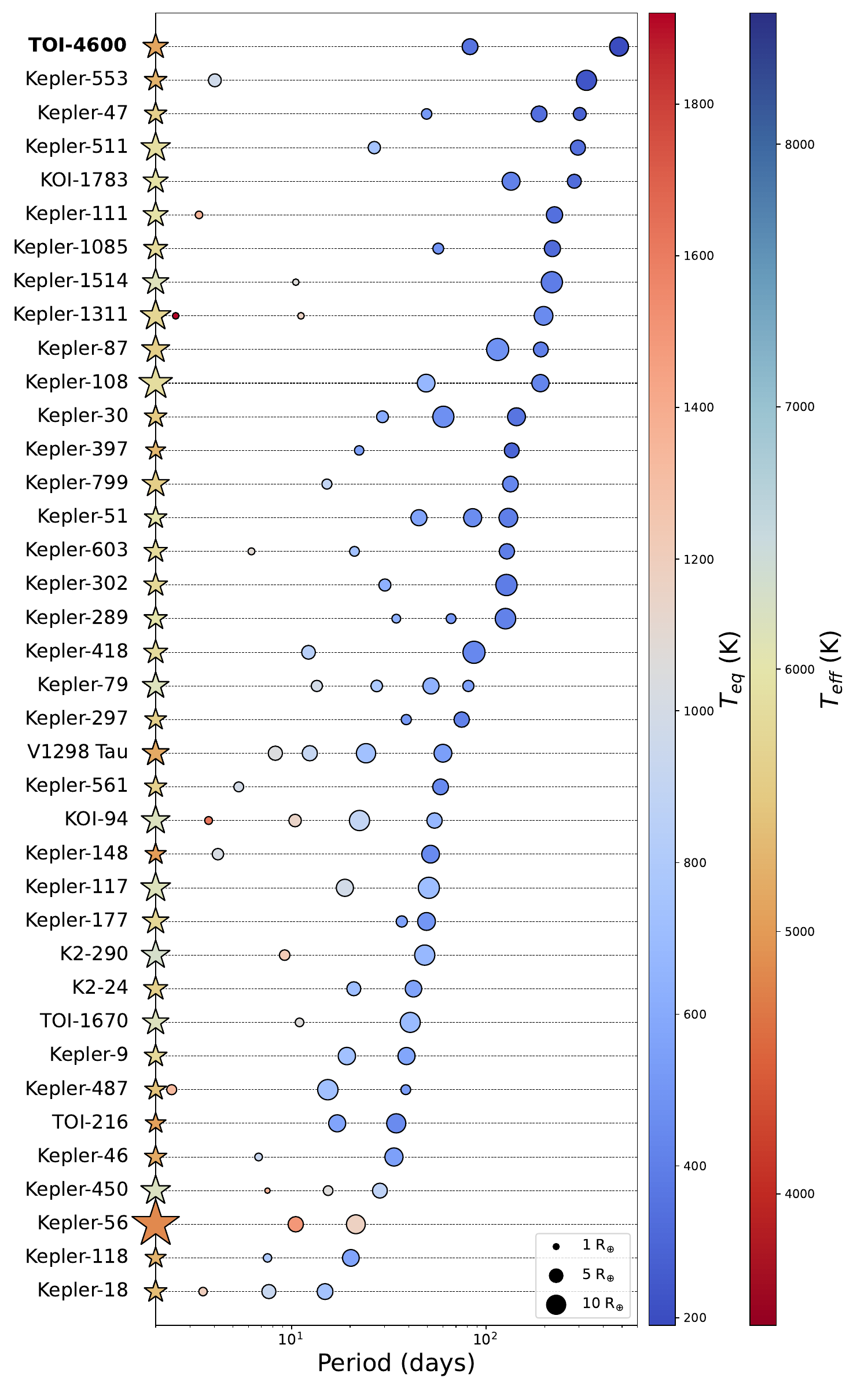}
\end{minipage}
\caption{{\it Top left:} Period-radius diagram of verified transiting planets orbiting stars with $V<13$ (black) and $V>13$ (gray), as of November 2022 (NASA Exoplanet Archive). {\it Bottom left:} Equilibrium temperature diagram (assuming albedo $a$ = 0) for the same sample. {\it Right:} The TOI-4600 system and other warm gas giant systems with multiple transiting planets. TOI-4600 c is the coldest and has the longest orbital period of any transiting planet in these systems.}
\label{fig:multi_per_teq_rad_comp}
\end{figure*}

Further characterization of the planet orbits with precise radial velocity measurements will reveal whether the orbits are actually eccentric or not, which in turn will shed light on the formation and evolution of this system. Various mechanisms have been proposed to explain the wide range in the eccentricities of warm Jupiters. Highly eccentric warm Jupiters can be explained by high-eccentricity migration models, including the Kozai cycle \citep{2007ApJ...669.1298F} and planet-planet scattering \citep{2014ApJ...786..101P}. Low eccentricity can be explained by a combination of disk migration \citep{1980ApJ...241..425G} and in-situ formation models, which result in well-aligned systems that often have smaller planets near the warm Jupiters \citep{2016ApJ...817L..17B}. A rarely-seen system with two such types of planets may play a significant role in advancing our understanding of planet formation and evolution. There are currently fewer than three dozen multi-transiting-planet systems with a warm Jupiter, and TOI-4600 c is the longest-period and coldest transiting planet in the sample (Figure \ref{fig:multi_per_teq_rad_comp}). 

TOI-4600 is well-suited for mass and orbit characterization with radial velocity measurements. It is currently the only system brighter than $V=13$ that hosts two long-period ($P>50$ d) transiting giant planets. Using conservative 3-$\sigma$ lower limit mass estimates from the empirical mass-radius relations from \citet{2017ApJ...834...17C} of $\sim 31$ and $\sim 48\, M_{\oplus}$, we predict RV semi-amplitudes of $\sim 5$ \ms\,  and $\sim 4$ \ms\,  for planets b and c, respectively. Given the lack of activity of the host star as indicated by the \tess\, photometry and TRES spectroscopy, mass and orbital characterization using RVs is well within the capabilities of current facilities. The wide separation between the star and inner planet and between the planets themselves may mean there are additional planets in the system, which could be detected with RV follow-up.

Additionally, the system is amenable to projected obliquity measurements through observations of the Rossiter-McLaughlin effect, with predicted semi-amplitudes of $\sim 12$ and $\sim 18$ \ms\, for planets b and c, respectively. As with the mass measurements, obliquity measurements for the system are also well within the reach of current facilities and would be among the longest-period planets with measured obliquities. However, we note that the long periods and transit durations would make this challenging, particularly for planet c.

While unlikely, TOI-4600 c may be detected in an upcoming Gaia data release using the astrometric technique. Given the system parameters and assuming a median mass estimate, the expected astrometric signature of the planet is approximately 1 $\mu$as. While this would be an order of magnitude below the detection threshold \citep{2014ApJ...797...14P}, if the planet mass is instead like Jupiter's or larger, it could produce a detectable signal.

With equilibrium temperatures of $\sim350$ K and $\sim190$ K, TOI-4600 b and c join the sparsely populated yet growing list of temperate/cool gas giant planets. Using median mass estimates from \citet{2017ApJ...834...17C} of $\sim 36$ and $\sim 63\, M_{\oplus}$ for planets b and c, respectively, we obtain Transit Spectroscopy Metric (TSM) values of $\sim 30$ for both planets. Among the nearly 400 transiting exoplanets with orbital periods greater than 50 days, fewer than two dozen have TSM values greater than 30 making TOI-4600 b and c two of the more promising targets for atmospheric characterization. Additionally, TOI-4600 is close to the northern continuous viewing zone for JWST \citep{2006SSRv..123..485G}. Transmission spectroscopy could be used to search for different molecules, such as methane, ammonia, carbon dioxide, and water \citep{2015ApJ...814..154D, 2020Natur.585..363V}, that could also be used to probe the history of the system \citep{2022ApJ...937...36P}.

\section{Summary} \label{sec:summary}

We use \tess\, observations and ground-based follow-up observations to statistically validate the planetary nature of two candidates orbiting the early K dwarf TOI-4600. TOI-4600 b is a sub-Saturn sized planet (6.80 $R_{\oplus}$) on a 82.69 d orbit. TOI-4600 c is a Saturn sized planet (9.42 $R_{\oplus}$) on a 482.82 d orbit. TOI-4600 c is the longest-period planet and among the coldest planets (in terms of equilibrium temperature for a specified albedo) discovered by \tess\, to date. As a system well-suited for radial velocity follow-up, the characterization of the masses and orbits of the two planets and search for additional planets will help inform formation and evolution theories of warm Jupiter systems. Additionally, both planets are excellent targets for the atmospheric characterization of warm and cool giant planets which have so far not been characterized. TOI-4600 highlights \tess's ability to not only detect long-period planets, but detect long-period planets that present great opportunities for further characterization.


\section*{Acknowledgements}
We thank the anonymous reviewer for their helpful comments, which have helped improve the paper. Funding for the TESS mission is provided by NASA's Science Mission Directorate. We acknowledge the use of public TESS data from pipelines at the TESS Science Office and at the TESS Science Processing Operations Center. Resources supporting this work were provided by the NASA High-End Computing (HEC) Program through the NASA Advanced Supercomputing (NAS) Division at Ames Research Center for the production of the SPOC data products. This research has made use of the Exoplanet Follow-up Observation Program website, which is operated by the California Institute of Technology, under contract with the National Aeronautics and Space Administration under the Exoplanet Exploration Program. This research has made use of the NASA Exoplanet Archive, which is operated by the California Institute of Technology, under contract with the National Aeronautics and Space Administration under the Exoplanet Exploration Program. This work makes use of observations from the LCOGT network. DD acknowledges support from the TESS Guest Investigator Program grants 80NSSC21K0108 and 80NSSC22K0185, and NASA Exoplanet Research Program grant 18-2XRP18\_2-0136. HPO’s and SU's contributions have been carried out within the framework of the NCCR PlanetS supported by the Swiss National Science Foundation under grants 51NF40\_182901 and 51NF40\_205606. The postdoctoral fellowship of KB is funded by F.R.S.-FNRS grant T.0109.20 and by the Francqui Foundation. KKM acknowledges support from the New York Community Trust Fund for Astrophysical Research. This work made use of \textsf{exoplanet} \citep{exoplanet:joss,exoplanet:zenodo} and its dependencies \citep{exoplanet:agol20,exoplanet:arviz, exoplanet:astropy13, exoplanet:astropy18, exoplanet:luger18,exoplanet:pymc3, exoplanet:theano, exoplanet:vaneylen19}.

Some of the data presented in this paper were obtained from the Mikulski Archive for Space Telescopes (MAST) at the Space Telescope Science Institute. The specific observations analyzed can be accessed via DOI:\dataset[10.17909/0cp4-2j79]{https://doi.org/10.17909/0cp4-2j79}, DOI:\dataset[10.17909/t9-nmc8-f686]{https://doi.org/10.17909/t9-nmc8-f686}, and DOI:\dataset[10.17909/t9-yk4w-zc73]{https://doi.org/10.17909/t9-yk4w-zc73}.

%

\vspace{5mm}
\facilities{\tess, LCOGT, Exoplanet Archive}


\software{AstroImageJ \citep{Collins:2017}, astropy \citep{2013A&A...558A..33A,2018AJ....156..123A}, TAPIR \citep{Jensen:2013}
          }





\bibliography{toi4600}{}
\bibliographystyle{aasjournal}




\appendix
\section{\texttt{allesfitter} Figures}
\restartappendixnumbering

\begin{figure*}[!h]
    \centering
    \includegraphics[width=\textwidth]{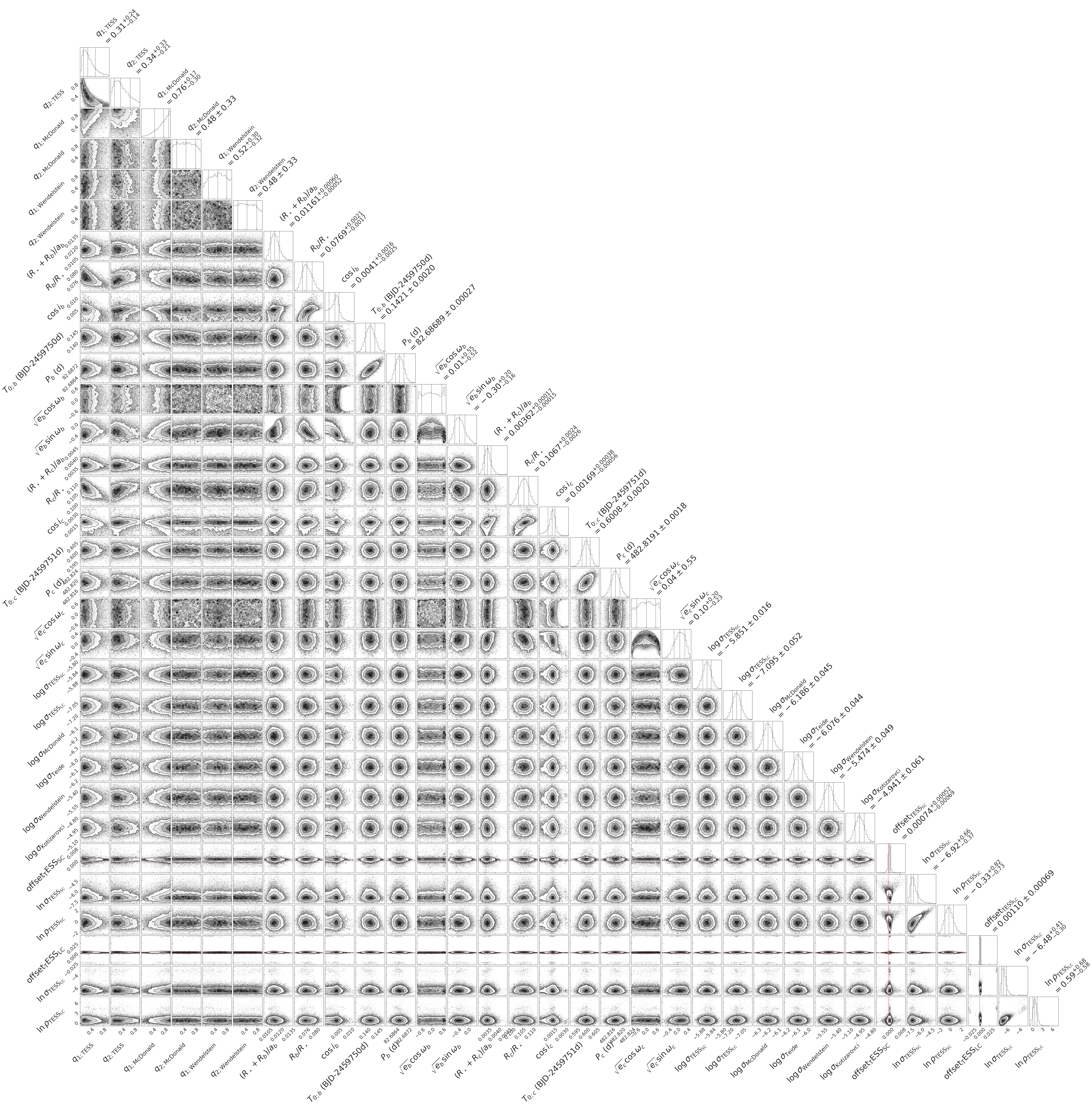}
    \caption{Corner plot of the fitted parameters obtained from \texttt{allesfitter}.}
    \label{fig:corner_fitted}
\end{figure*}

\begin{figure*}[!h]
    \centering
    \includegraphics[width=\textwidth]{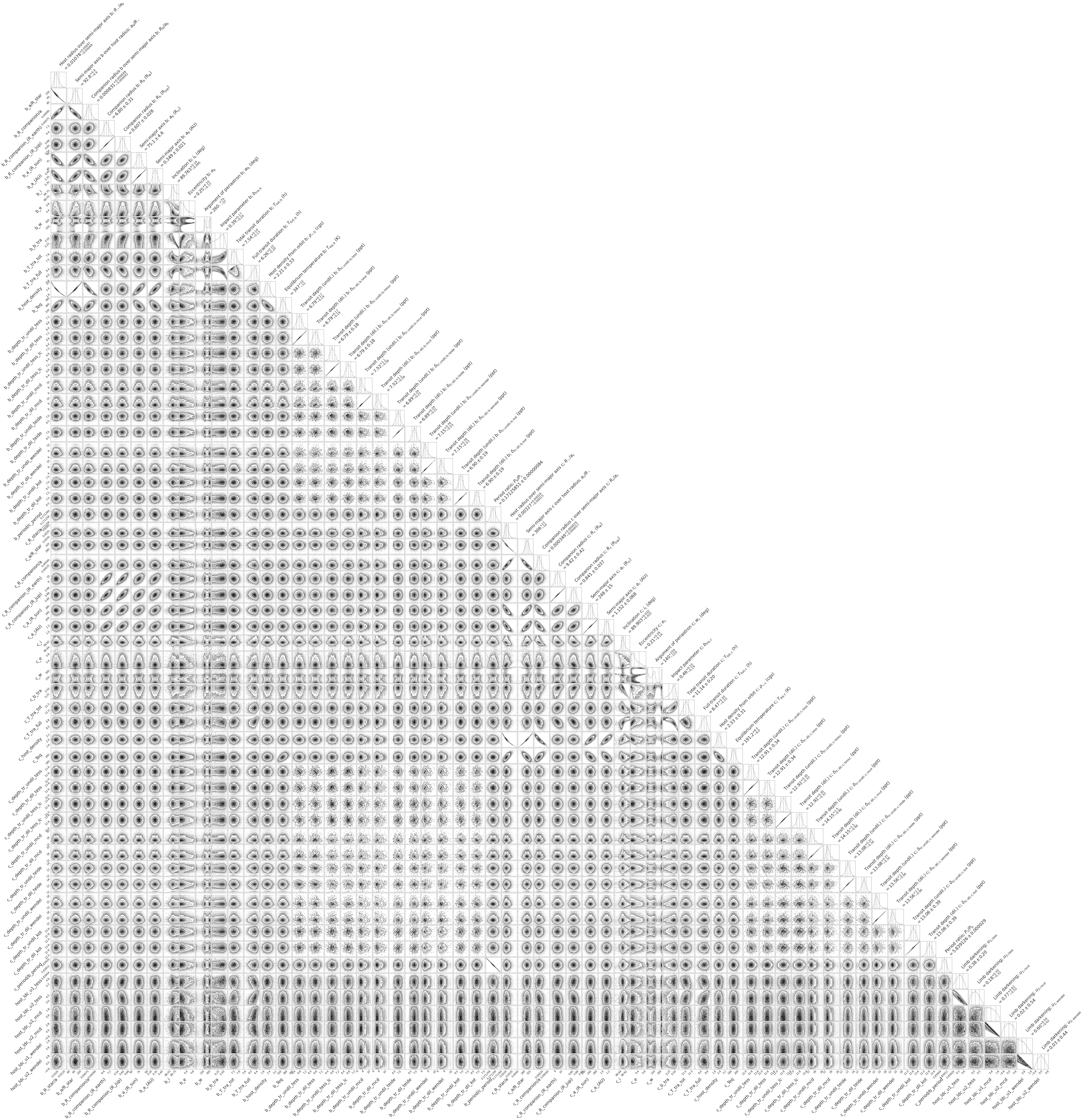}
    \caption{Corner plot of the derived parameters obtained from \texttt{allesfitter}.}
    \label{fig:corner_fitted_der}
\end{figure*}


\end{document}